\documentclass{aastex631}

\usepackage{booktabs}
\usepackage{amsmath}
\usepackage{subfigure}
\usepackage{cancel}
\begin{document}

\title{Jet Lorentz factor constraint for GRB 221009A based on the optical depth of the TeV photons}

\correspondingauthor{Yuan-Chuan Zou}
\email{zouyc@hust.edu.cn}

\author[0009-0007-0866-4265]{Duan-yuan Gao}
\affiliation{School of Physics, Huazhong University of Science and Technology, Wuhan 430074, China}

\author[0000-0002-5400-3261]{Yuan-Chuan Zou}
\affiliation{School of Physics, Huazhong University of Science and Technology, Wuhan 430074, China}



\begin{abstract}
The recent detection of tera-electronvolt (TeV) photons from the record-breaking gamma-ray burst GRB 221009A during its prompt phase poses challenges for constraining its Lorentz factor. We re-evaluate the constraints on the jet Lorentz factor considering a two-zone model, wherein the TeV photons originate from the external shock region while the lower energy MeV photons come from the internal prompt emission region. By properly accounting for the evolution of the MeV photon spectrum and light curve, we calculate the optical depth for TeV photons and derive a minimum Lorentz factor about 300. 
It is consistent with the afterglow modeling for the TeV emission.
\end{abstract}

\keywords{Gamma-ray bursts(629) --- High energy astrophysics (739) --- Relativistic jets(1390) }


\section{Introduction} \label{sec:intro}
GRB 221009A is the brightest gamma-ray burst ever observed in recorded history. Existing statistical data suggest that such an event occurs, on average, once every ten thousand years \citep{2023ApJ...946L..31B}. 
The central engine of GRB 221009A is very close to Earth, with a redshift of 0.151 \citep{2022GCN.32648....1D} and trigger time ($T_{\rm 0}$) of 2022 October 9 at 13:16:59.99 UTC\citep{2023arXiv230314172L}. The prompt burst's isotropic energy is exceptionally high, on the order of $10^{55} \,\rm{erg}$, concentrated in the MeV band \citep{2023arXiv230301203A,2023arXiv230314172L,2023ApJ...949L...7F}. This indicates that the main burst emitted a significant amount of MeV photons, allowing for spectral fitting in different time intervals. 
\cite{2023arXiv230301203A,2023arXiv230314172L} have all performed band power-law spectral fitting for different time intervals of GRB 221009A. 
Approximately 226 seconds after the burst, \cite{2023Sci...380.1390.} detected 64,000 TeV photons, including a photon with an energy of 7 TeV. 
Some researchers propose that the origin of TeV photons may be attributed to certain hadronic processes in the prompt burst \citep{2023SCPMA..6689511W}. 
Conversely, others believed that TeV photons may originate from external Inverse Compton and proton synchrotron radiation in the external shock \citep{2023ApJ...947L..14Z}. 
However, \cite{2023Sci...380.1390.} argued that these photons originate from synchrotron self-Compton (SSC) emission in the afterglow.
The observation of very high-energy TeV emission during the prompt phase suggests that the Lorentz factor might have reached an extremely large value, which may raise certain concerns regarding the self-consistency of the overall picture.

Lorentz factor is always an important property of GRB jets. 
As discussed in \citet{2021ApJ...909L...3Z}, the bulk Lorentz factor in the prompt phase and in the deceleration phase are different.
Various methods have been proposed to either directly obtain or constrain the Lorentz factor of the GRB jet at different phases.
In the external shock scenario, the deceleration time of the material, reflected in the peak of the light curve, can be used to estimate the Lorentz factor \citep{1997ApJ...476..232M,1999ApJ...517L.109S,1999ApJ...520..641S,2000ApJ...545..807K}. 
By comparing X-ray and optical emissions and fitting the parameters of the emitting region, the Lorentz factor can also be estimated \citep{1999ApJ...517L.109S,2003ApJ...595..950Z}. 
Considering the prompt emission should not be Compton scattered by the electrons associated with ejected baryons, one could set a lower limit \citep[Limit C in][]{2001ApJ...555..540L}, which is generally less than one and is useful only for strong GRBs.
Additionally, the Lorentz factor can be estimated through the observation of the thermal component \citep{2005ApJ...635..516N,2007ApJ...664L...1P,2015ApJ...800L..23Z}. 
Another approach involves considering that the  high-energy gamma-ray photons to escape without being absorbed by lower-energy gamma-ray photons. 
Two models have been proposed for this: the single zone model, where high-energy and low-energy photons originate from the same region \citep{2001ApJ...555..540L}, and the two-zone model, which suggests that high-energy and low-energy photons come from different regions \citep{2011ApJ...726L...2Z}. 
Most high-energy emissions have shown smooth light curves, consistent with the afterglow model \citep{2010MNRAS.403..926G}. 
For GRB 221009A, the smooth TeV light curve indicates an external shock origin. \citet{2023Sci...380.1390.} have successfully modeled the entire light curve and spectrum with a standard afterglow plus SSC model. 
This suggests that the TeV emission should not originate from the same internal shock regions that produce the prompt MeV emission.
Therefore, the constraint on the Lorentz factor should consider the two-zone effect.

For GRB 221009A, \cite{2023arXiv230314172L} estimated the minimum values of the Lorentz factors at different stages. During the precursor radiation phase, the minimum bulk Lorentz factor was found to be 188. In the main burst phase, considering that the emission site must be optically thin, the minimum bulk Lorentz factor for the minimum variability timescale ($t_{\rm{var}} = 0.1 \rm{s}$) was calculated to be 1040. According to the single zone model, the minimum Lorentz factor estimated for the highest-energy photon at 99.3 GeV was 1560, and considering a more relativistic situation could lower it to 780. Moving on to the afterglow phase, considering the interstellar medium model, the minimum Lorentz factor was determined to be 260, while in the stellar wind model, it was 282. Based on support from the light curve, \cite{2023Sci...380.1390.} adopted the uniform interstellar medium model to estimate the bulk Lorentz factor of the afterglow as 560. \cite{2023ApJ...947L..11Y} used data from the Fermi Gamma-ray Burst Monitor (precursor) and SATech-01/GECAM-C (main emission and flare) to fit the Lorentz factor of GRB 221009A using the SSC model in different time segments. The Lorentz factor was estimated to be around 250, reaching a maximum of 741.

In this work, we check the constraint by the $\gamma \gamma$ absorption on the TeV photons.
As the light curve of GRB 221009A has smooth temporal profile which supports the TeV photons come from external shock, we mainly focus on the so called two-zone model.
In section \ref{sec:re}, we introduce the single zone model and the two-zone model considering spectral and variability corrections, as well as present the simplified equations for the two-zone model. 
In section \ref{sec:results}, we calculate the optical depth for GRB 221009A for fixed Lorentz factors and minimum Lorentz factors. 
Finally, in the section \ref{sec:con}, we provide conclusions and future prospects.

\section{restriction to Lorentz factor}
\label{sec:re}
\subsection{Single zone model}
If TeV photons and MeV photons are produced in the same location, they may collide to generate electron pairs. This leads to a limit for the minimal Lorentz factor of single zone model \citep{2001ApJ...555..540L,2008MNRAS.384L..11G}.
The energy required for photon annihilation is much lower than the peak energy of the spectrum. Therefore, we treat the spectrum as a single power-law spectrum, using the photon index in the low-energy range. 
So, the photon spectrum of MeV photons (i.e., the number density of photons per unit time per unit area per unit energy) can be represented as $fE^{-\beta}$, where f is the normalization factor, and $E$ is the photon energy.
The Optical depth of single power-law spectrum was described in \cite{2008MNRAS.384L..11G} as case 1, i.e., equations (13) and (14). By replacing $\beta_2$ with the low-energy index $\beta$, changing the optical depth $\tau_{int}$ to $\tau$, and replacing the maximum energy of electrons $E_{\gamma h}^o$ with $E_{\max}$, we obtain:
\begin{equation}
 \tau = \frac{A_1(E_{\max})}{R^2} \left(\frac{\Gamma}{1+z}\right)^{-2\beta+2},
 \label{single_gamma}
\end{equation}
where $R$ is the emitting radius from the central engine, which is different in different model, $z$ is the redshift, $\beta = 1.69$  is the photon spectral index for GRB 221009A \citep{2023arXiv230314172L}, $E_{\max} = \rm{7\ TeV}$, and $A_1$ is defined as 
\begin{eqnarray}
\label{eq:single_tau}
 A_1 &\equiv& \frac{(11/180)\sigma_{\rm{T}} f d_z^2 R}{(\beta-1) \Gamma^2 c }\left(\frac{E_{\max}}{m_{\rm{e}}^2 c^4}\right)^{\beta-1} ,\nonumber \\
\end{eqnarray}
where $\sigma_{\rm T}$ is the Thomson cross section, $m_{\rm e}$ is the rest mass of an electron, $c$ is the speed of light, $d_z$ is the luminosity distance. 
As the photon with energy $E_{\max}$ has been observed, it indicates the optical depth $\tau < 1$, which derives a minimum Lorentz factor from equation (\ref{single_gamma}).
Since what we have observed is the photon flux per second rather than the total flux, for convenience, we replace $f_1^o$ of equation (14) in \cite{2008MNRAS.384L..11G} with $fR/(\Gamma^2 c)$, and $f d_z^2 = 3.32 \times {10}^{54} {\rm erg}^{\beta-1} {\rm s}^{-1} $ which can be derived from the isotropic luminosity and peak energy.
To be consistent with \cite{2001ApJ...555..540L}, we change the coefficient $3/[8(\beta_1^2-1)]$ in equation (14) of \cite{2008MNRAS.384L..11G} to $11/[180(\beta - 1)]$.
We set the optical depth equal to 1 to obtain the minimum Lorentz factor.
In the case of internal shocks, emitting radius is $R = 2 \Gamma^2 c t_v/(1+z)$, where $t_v = 0.1{\rm s}$ \citep{2023arXiv230314172L} is the variability time scale of the prompt emission.
We finally get $\Gamma_{\min} = 2164$.
This value is much larger than that given by  \cite{2023arXiv230314172L} ($\Gamma_{\min} = 1560$), because the maximum energy of the high energy photon was 99.3 GeV in {\it Fermi}-LAT data.
This is far greater than the estimated value by the afterglow model, which is unreasonable.

If we adopt the Internal Collision-Induced Magnetic Reconnection and Turbulence (ICMART) model, the radius can reach to a larger value of approximately $R = 2\Gamma^2 c t_v/(1+z)$\citep{2011ApJ...726...90Z}, where $t_v \sim 20 {\rm s}$ represents the characteristic timescale of the long pulse. In this case, the minimum Lorentz factor can be reduced to 808.

\subsection{Two-zone model}
From the TeV light curve given by \cite{2023Sci...380.1390.}, it can be well modeled by the external shock. Therefore, we turn to consider the two-zone model \citep{2011ApJ...726L...2Z}. We assume that TeV photons come from the external shock, while MeV photons originate from the prompt internal shocks, which is consistent with \cite{2023Sci...380.1390.}. 

In the two-zone model, we assume that MeV and TeV photons come from distinct regions. 
The MeV photons come from an emitting region with a radius $R_{\rm{M}}$, a Lorentz factor of $\Gamma_{\rm{M}}$, and an angular width of $\theta_{\rm{M,J}}$. 
Similarly, the TeV photons come from a different emitting region with a radius $R_{\rm{T}}$, a Lorentz factor of $\Gamma_{\rm{T}}$, and an angular width of $\theta_{\rm{T,J}}$.
Notice the notation of T (meaning TeV) was used as G (meaning GeV) in \cite{2011ApJ...726L...2Z}.
To get a more accurate constraint, we extend the T90 in \cite{2011ApJ...726L...2Z} to a real time varying MeV light curve.

\begin{figure}[ht!]
\plotone{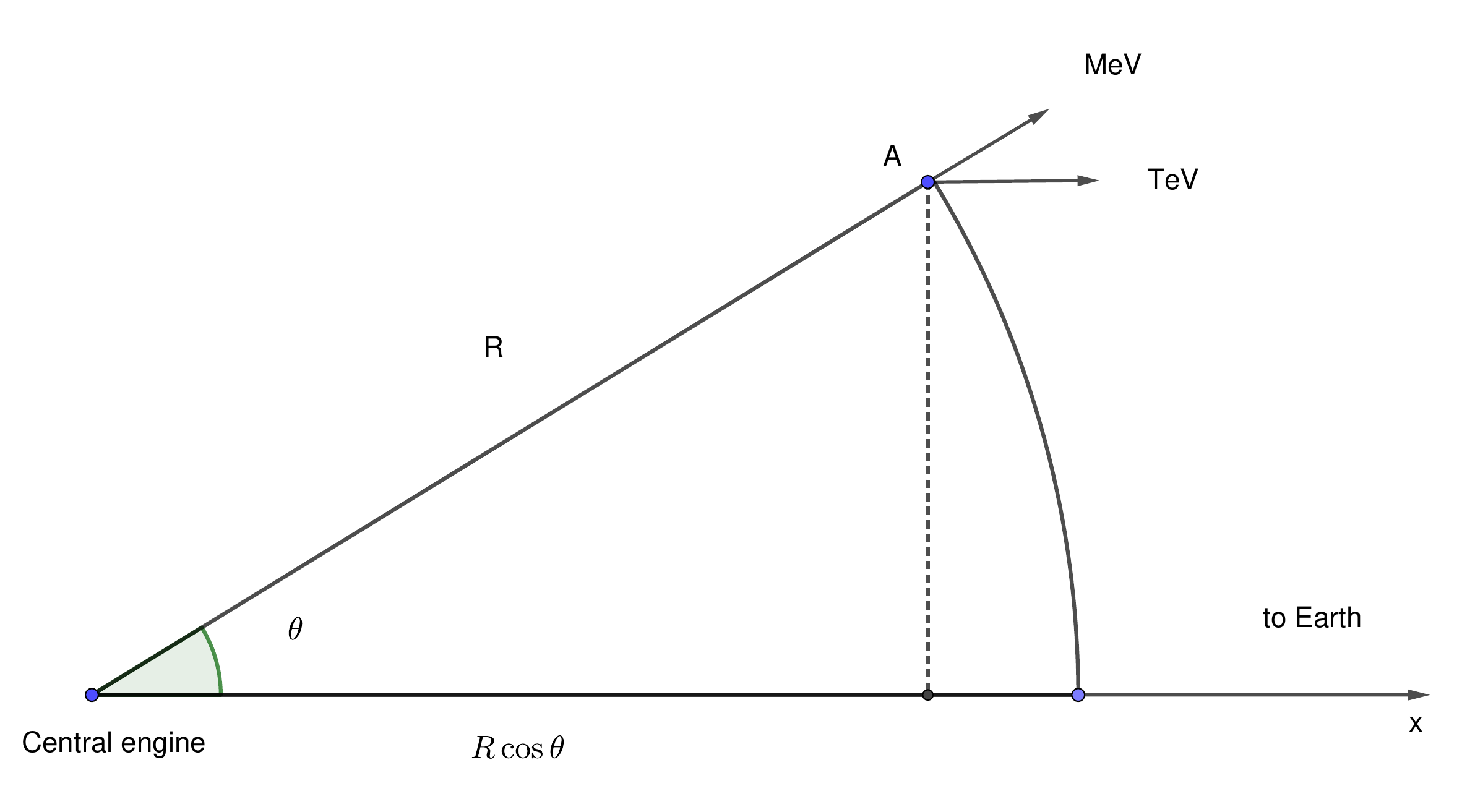}
\caption{A schematic diagram illustrates the geometry of the difference in retarded time between MeV photons and TeV photons. When TeV photons collide with MeV photons at point A. 
the MeV photons will not reach Earth as the direction is not pointing to the Earth, but we can determine their number density from the photons arriving at Earth with the same retarded time.
The MeV photons move radially at $R$, while the TeV photons move along the x-axis at $x = R\cos\theta$ when they are colliding.}
\label{fig:general3}
\end{figure}

  Figure \ref{fig:general3} shows a sketch for the geometry of TeV and MeV photons collision and propagating. A more detailed illustration can be seen in \cite{2011ApJ...726L...2Z}.
We set the x-axis toward Earth, and the zero point of x is at the central engine.

To describe the distribution of MeV photons, we utilize the photon field, assuming that the MeV photon comes from a radius $R_{\rm M}$, which is much smaller than R. 
In other words, we consider the MeV photons as originating from the center approximately. 
Hence, the photon number density is decay with $R^2$ , in the other word, the photon number density times radius square only related to the radius $R$ and time $t$, denoted as $R^2 n_{\rm M}(R,t)$. 
Since photons always move at the speed of light, we can express $R^2 n_{\rm M}$ as a function of the retarded time $t_{\rm R,M} = t - R/c$. 
The retarded time does not change with the movement of photons.
Taking the retarded time obtained from the trigger time and the position of Earth as the zero point, the retarded time is equivalent to the time on the light curve.
Similarly, for TeV photons that come to Earth, retarded time is $t_{\rm R,T} = t - x/c$.

Considering a TeV photon at $(R, \theta)$ that moves toward Earth and collides with a MeV photon, the time $t$ of the TeV photon and the MeV photon is the same. So the difference in their retarded time is:
\begin{equation}
 \delta t_{\rm{R}} = t_{\rm R,T} - t_{\rm R,M} = (1+z) \left(\frac{R}{c} - \frac{x}{c}\right)  =  (1+z)\frac{R}{c}(1-\cos\theta)  \approx \frac{1}{2}(1+z)\theta^2 \frac{R}{c}.
  \label{eq:tm}
\end{equation}

Due to the TeV photons direction being parallel to the line connecting the central engine and the Earth, the collision angle between TeV photons and MeV photons is $\theta$, and the distance of the TeV photons to the x-axis remains constant. This means:
\begin{equation}
    R \sin\theta = R_{\rm{T}} \sin\theta_{\rm{T}} = {\rm const.} \approx R \theta. 
\end{equation}
Combining equation \eqref{eq:tm}, we can deduce that $\delta t_{\rm{R}} \propto \frac{1}{R}$. Therefore, at an infinite distance, $\delta t_{\rm{R}}=0$, indicating that the MeV photons and TeV photons that collide at Earth will reach Earth simultaneously. This natural consequence confirms the consistency in the timing of their arrival at Earth.

Because $n_{\rm M}$ is zero when $t_{\rm{R}}>T_{90}$, so we can only consider $R$ smaller than a certain distance \citep{2011ApJ...726L...2Z}:
\begin{equation}
R_{\max}(R_{\rm{T}},\theta_{\rm{T}}) =  \frac{R_{\rm{T}}}{1-2 c T_{\rm{90}}/[(1+z)R_{\rm{T}} \theta_{\rm{T}}^2]}.
\end{equation}
 Due to the high brightness of GRB 221009A, the spectrum of MeV photons is very accurate, even in each time interval. Therefore, we can describe the spectrum using the Band function, with the number density $n_{\rm{0,MeV}}$, the power law photon indices $\alpha$ and $\beta$, and peak energy $E_{\rm{P}}$ depend on time: 
\begin{eqnarray}
 n_{\rm{MeV}}(E_{\rm{M}}) \simeq n_{\rm{0},\rm{MeV}} (R,t_{\rm{R}})  \left\{  
  \begin{array}{ll}
   \left[\frac{E_{\rm{M}}}{E_{\rm{P}}(t_{\rm{R}})}\right]^{-\alpha_{\rm{M}}(t_{\rm{R}})}, \phantom{00} & E<E_{\rm{P}}(t_{\rm{R}}) , \\
   \left[\frac{E_{\rm{M}}}{E_{\rm{P}}(t_{\rm{R}})}\right]^{-\beta_{\rm{M}}(t_{\rm{R}})}, \phantom{00} & E_{\rm{p}}(t_{\rm{R}})<  E < E_{{\rm{M}},{\max}},
  \end{array}
   \right. 
\end{eqnarray} 
where $\alpha_{\rm{M}}(t_{\rm{R)}}$ is the low energy photon index and $\beta_{\rm{M}}(t_{\rm{R}})$ is the high energy photon index. We aim to use the spectral indices to obtain a more accurate number density of the MeV photons. Considering the high-energy power law spectral index $\beta$ is approximately 2, and the upper energy boundary of HXMT is close to $E_{\rm{P}}$, we need to consider the upper limit of the integral. The number density of the MeV photons can be expressed as follows:
\begin{equation}
    n_{0,\rm{Mev}} = \frac{L_{\rm{M}}(t_{\rm{R}})}{(1+z)^2 4\pi R^2c \int_{0}^{E_{\max,\rm{o}}} E \left(\frac{E}{E_{\rm{P}}(t_{\rm{R}})}\right)^{-{\alpha}} {\rm d}E } = \frac{L_{\rm{M}}(t_{\rm{R}})}{(1+z)^2 4\pi R^2c {E_{\rm{P}}^{*2}(t_{\rm{R}})}},
    \label{eq:n0}
\end{equation}
where $E_{\max,\rm{o}}$ is the maximal energy of the observed MeV photon, the power law photon index $\alpha$ can take either $\alpha_{\rm{M}}$ or $\beta_{\rm{M}}$ at different $E$, and $E_p^*(t_{\rm{R}})$ is the effective peak energy:
\begin{equation}
    E_p^*(t_{\rm{R}})=E_p(t_{\rm{R}}) \left[\frac{1}{2-{\alpha(t_{\rm{R}})}}-\frac{1}{\beta(t_{\rm{R}})-2}\left(1+E_{\rm{P}}^{\beta(t_{\rm{R}})-2}E_{\max}^{2-\beta(t_{\rm{R}})}\right)\right]^{{1}/{2}}.
    \label{eq:epstar}
\end{equation}

The interaction between a TeV photon and a MeV photon can occur as long as the total energy in the center-of-mass system is larger than $2 m_{\rm{e}} c^2$. Therefore, only photons that are more total energetic than $E_{M,\min}$ can annihilate a TeV photon with energy $E_{\rm{T}}$: 
\begin{equation}
E_{M,\min}  = \frac{ 2 (m_{\rm{e}} c^2)^2 }{(1+z)^2 E_{\rm T} (1-\cos \theta)} \simeq \frac{4(m_{\rm{e}} c^2)^2 }{(1+z)^2 \theta^2 E_{\rm T}} . 
\label{eq:e_min}
\end{equation}
For MeV photons that are sufficiently energetic, the cross-section depends on the energy as follows:
\begin{equation}
 \sigma = \frac{3}{16} \sigma_{\rm{T}} (1-\hat \beta ^2) \left[(2-\hat \beta^2)\ln \frac{1+\hat \beta}{1-\hat\beta}-2\hat\beta (2-\hat\beta^2)\right],
 \label{eq:sigma}
\end{equation}
where 
\begin{equation}
\hat \beta = \sqrt{1-2\frac{m_{\rm{e}} c^2}{(1+z)E_{\rm{M}}} \frac{m_{\rm{e}} c^2}{(1+z)E_{\rm{T}}} (1-\cos\theta)^{-1}},
\label{eq:beta}
\end{equation}
$\hat \gamma = \sqrt{1-\hat \beta^{2}}$ 
and $\sigma_{\rm{T}} \simeq 6.65\times 10^{-25} \mathrm{cm^2}$ is the Thompson cross-section.

Now, we can estimate the overall optical depth for the interaction of a TeV photon with the MeV pulse:
\begin{equation}
\tau(\theta_{\rm{T}},E_{\rm{T}}) = \int_{R_{\rm{T}}}^{R_{\max}(R_{\rm{T}},\theta_{\rm{T}})} {\rm d}R  \int_{E_{M,\min}}^{E_{M,\max}} {\rm d}E_{\rm{M}}  \frac{{\rm d} ^3 n_{\rm{MeV}}(t_{\rm{R}})}{{\rm d}E_{\rm{M}} {\rm d}\theta_{\rm{M}}} \sigma (1-\cos \theta),
\label{eq:tau_theta}
\end{equation}
where $E_{\max}$ is the upper limit of MeV component.The effective optical depth for a photon observed with an energy $E_{\rm{T}}$ is  averaged over all angles:
\begin{equation}
e^{-\bar{\tau} (E_{\rm{T}})} = \frac{\int_{0}^{\theta_{\rm T,j}}  \mathcal{D}^{-(3+\beta_{\rm{T}})} e^{-\tau(\theta_{\rm{T}},E_{\rm{T}})}     \theta_{\rm{T}} {\rm d}\theta_{\rm{T}}}{ \int_{0}^{\theta_{\rm T,j}} \mathcal{D} ^{-(3+\beta_{\rm{T}})}    \theta_{\rm{T}} {\rm d}\theta_{\rm{T}}},
\label{eq:e_tau_complete}
\end{equation}
where $\mathcal{D}=\Gamma_{\rm{T}}({1-\beta_{\rm bulk,T} \cos \theta_{\rm{T}}})$ is the Doppler factor, $\beta_{\rm bulk,T}=(1-1/\Gamma_{\rm{T}}^2)$ is the bulk velocity in $c$, and $\beta_{\rm{T}}$ is the photon index of the TeV emission. 
Since LHAASO observes a maximum angle of 0.8° corresponding to the afterglow, which is much larger than $1/\Gamma_{\rm{T}}$, and due to the beaming effect, we cannot see photons beyond $1/\Gamma_{\rm{T}}$. Therefore, our integration upper limit is $1/\Gamma_{\rm{T}}$.

By setting the effective optical depth $\bar{\tau}$ to 1, one can solve equation (\ref{eq:e_tau_complete}) to get $\Gamma_{\rm T, \min}$.

\subsection{A simplified model}
To improve the feasibility of our model, we make some approximations to derive an analytic formula. We have found some mistakes in \cite{2011ApJ...726L...2Z}, and we will correct and address them in the subsequent formula descriptions. We will later compare these results with the full formula. First, since for a specific $\theta_{\rm{T}}$, only one power-law index of the spectrum determines the optical depth, we use a single power-law to describe the spectrum. Second, we approximate the cross-section as $\sigma_{\rm{T}}/ (3\hat{\gamma}^2)$. Third, we assume that the angle of MeV photons is larger than the angle of TeV photons. We define $l\equiv \frac{R}{R_{\rm{T}}}$ and the maximum radius $l_{\max} \equiv R_{\max}/R_{\rm{T}} = 1 /[1-2 \Delta_{M} / (R_{\rm{T}}\theta_{\rm{T}}^2)]$. The minimal energy of the MeV photon for pair production is \citep{2011ApJ...726L...2Z}: 
 \begin{equation}
      E_{min}\approx \frac{4(m_{\rm{e}}c^2)^2}{(1+z)^2E_{\rm{T}} \theta_{\rm{T}}^2}l^2.
 \end{equation}
 The number density of the MeV photons is:
 \begin{equation}
      N(E_{\rm{M}})\approx n_{0,\rm{Mev}}\left(E_{\rm{M}}\right)=\frac{(1+z)^{2{\alpha}-2}L_{\rm{M}}E_{\rm{T}}^{\alpha} E_{\rm{P}}^{\alpha} \theta_{\rm{T}}^{2{\alpha}}}{4\pi R_{\rm{T}}^2 c (2m_{\rm{e}}c^2)^{2{\alpha}}{E_{\rm{P}}^*}^2}l^{-(2{\alpha}+2)}y^{-{\alpha}},
      \label{eq:N_ap}
 \end{equation}
where $y=E_{\rm{M}}/E_{\rm{M,\min}}=\hat{\gamma}^2$. We have corrected $\hat{\gamma}$ (\cite{2011ApJ...726L...2Z}) to $\hat{\gamma}^2$, and also have removed the unnecessary $\pi$ in front of $m_{\rm{e}} c^2$ in the denominator of equation (\ref{eq:N_ap}) \citep[equation (12) in][]{2011ApJ...726L...2Z}. 
Considering the simplified equations above, we have the optical depth for a TeV photon emitted at $\theta_{\rm T}$:
\begin{equation}
    \begin{split}
         \tau(\theta_{\rm{T}}) & =\int_{R_{\rm{T}}}^{R_{\max}(R_{\rm{T}},\theta_{\rm{T}})}{\rm d}R\int_{E_{\rm{{M,\min}}}(\theta,R)}^{E_{\rm{{M,\max}}}}{\rm d}E_{\rm{M}}\cdot N(E_{\rm{M}})\times(1-\cos\theta)\frac{\sigma_{\rm{T}}}{3\hat{\gamma}^2} \\ 
         &\approx \frac{(1+z)^{2{\alpha}-4}L_{\rm{M}}\epsilon_{\rm{P}}^{\alpha} \epsilon_{\rm{T}}^{{\alpha}-1} \sigma_{\rm{T}}\theta_{\rm{T}}^{2{\alpha}}}{6\times 4^{\alpha_{\rm{M}}}\pi R_{\rm{T}} m_{\rm{e}}c^3{\epsilon_{\rm{P}}^*}^2{\alpha}}
         \left[\frac{1}{2{\alpha}+1}\left(1-l_{\max}^{-(2{\alpha}+1)}\right)-\left(\frac{1}{4}(1+z)^2\epsilon_{\max}\epsilon_{\rm{T}}\theta_{\rm{T}}^2\right)^{-{\alpha}}\times \left(1-\frac{1}{l_{\max}}\right)\right],   
    \end{split}     
    \label{eq:simple_tua}
\end{equation}
where $\epsilon_{\rm{P}} \equiv E_{\rm{P}}/m_{\rm{e}}c^2$, ${\epsilon}_{\rm{P}}^* \equiv {E}_{\rm{P}}^*/m_{\rm{e}}c^2$, $\epsilon_{\rm{T}}\equiv E_{\rm{T}}/m_{\rm{e}}c^2$ and $\epsilon_{\max}\equiv E_{\rm M,\max}/m_{\rm{e}}c^2$. We have corrected the power law index of (1+z) in the numerator from $2\alpha-3$ \citep[equation (13) in][]{2011ApJ...726L...2Z} to $2\alpha - 4$, which was a typo.

From this equation, we can see that $\tau$ is proportional to ${\theta_{\rm{T}}}^{2{\alpha}}$. By substituting into equation (\ref{eq:e_tau_complete}), the other two terms in the numerator of the integrated function are also increasing functions of $\theta_{\rm{T}}$, so we understand that the effective optical depth is mainly determined by the maximum value of $\theta_{\rm{T}}$. For GRB 221009A, the outflow angle is larger than $1/\Gamma_{\rm{T}}$, so the maximum value of $\theta_{\rm{T}}$ is $1/\Gamma_{\rm{T}}$. Since we believe the TeV photons come from the outflow, $R_{\rm{T}}$ can be expressed as $R_{\rm{T}}=2\Gamma_{\rm{T}}^2 c t_{\rm{T}}/(1+z)$. Finally, we obtain the minimum Lorentz factor:
\begin{equation}
    \begin{split}
        \Gamma_{\rm T,min} & \approx \left[\frac{(1+z)^{2{\alpha}-3}L_{\rm{M}}\epsilon_{\rm{T}}^{{\alpha}-1} \epsilon_{\rm{P}}^{\alpha} \sigma_{\rm{T}}}{12\times 4^{{\alpha}}\pi m_{\rm{e}}c^4  t_{\rm{T}}{\epsilon_{\rm{P}}^*}^2 {\alpha}}\right]^{\frac{1}{2{\alpha}+2}} \\
        & \approx 34 \left(\frac{30\times 10^{4 {\alpha}}}{{\alpha} \times 34^{2{\alpha}+2}}\right)^{\frac{1}{2{\alpha}+2}}\left[\left(\frac{1+z}{2}\right)^{2{\alpha}-3} L_{M,51}\epsilon_{\rm{T}}^{{\alpha}-1} \epsilon_{\rm{P}}^{\alpha}{\epsilon_{\rm{P}}^*}^{-2} t_{\rm{T,2}}^{-1}\right]^{\frac{1}{2{\alpha}+2}}.
    \end{split}
    \label{eq:simple_gamma}
\end{equation}

\section{results}
\label{sec:results}

\begin{figure}[ht!]
\plotone{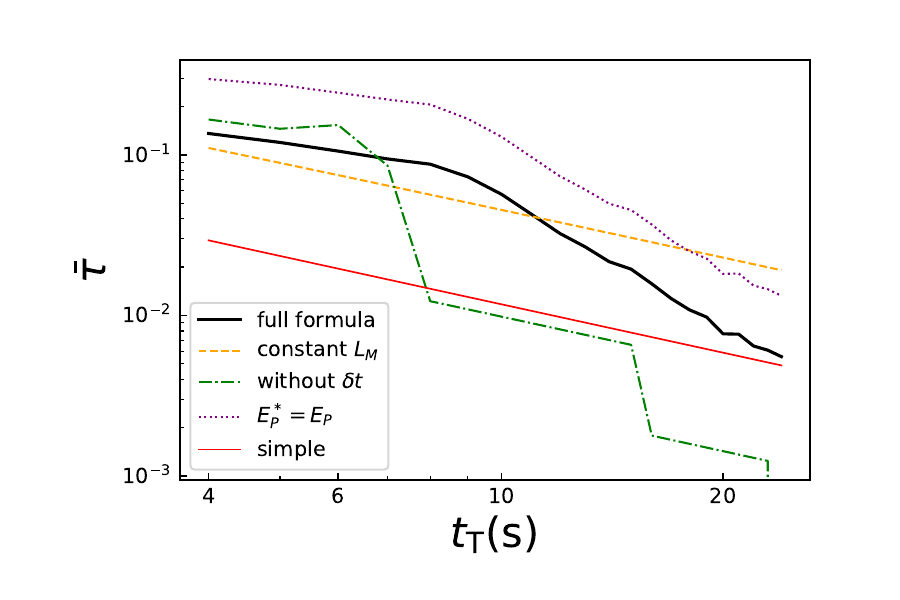}
\caption{The plot shows the optical depth for a photon with $E_{\rm{T}} = 7$ TeV as a function of $t_{\rm{T}}$ when the Lorentz factor $\Gamma_{\rm{T}} = 560$. Different lines in different colors and line styles represent the optical depth as a function of $t_{\rm{T}}$ for different methods.
The black thick line represents the result obtained using the full formula, i.e., equation (\ref{eq:e_tau_complete}). This is the most comprehensive and accurate calculation taking into account all relevant factors.
The orange dashed line is obtained by neglecting changes in luminosity, power-law indices, and peak energy. In this method, we substitute the variables in equation (\ref{eq:n0}) with their time-averaged values, while using equation (\ref{eq:e_tau_complete}) for other parts of the calculation.
The green dashed-dotted line considers only changes in $t_{\rm{T}}$ and neglects $\delta t_{\rm{R}}$, i.e., $\delta t_{\rm{R}}=0$, instead of using the full expression from equation (\ref{eq:tm}). Other parts of the calculation use equation (\ref{eq:e_tau_complete}).
The purple dotted line omits the correction for the photon number density with the power-law index of the spectrum, i.e., $E_{\rm{P}}^* = E_{\rm{P}}$, instead of using the full expression from equation (\ref{eq:epstar}). Other parts of the calculation use equation (\ref{eq:e_tau_complete}).
The red thin solid line is obtained using a simpler method, i.e., equation (\ref{eq:simple_tua}). 
}
\label{fig:general}
\end{figure}

Taking all the observational quantities into equation (\ref{eq:e_tau_complete}) and the related equations, we can finally calculate the average optical depth for a TeV photon with certain energy and at a certain observing time.
Figure \ref{fig:general} illustrates the dependence of $\bar{\tau}$ on $t_{\rm{T}}$ for a set of parameters of GRB 221009A: $z = 0.151$ \citep{2022GCN.32648....1D}, $E_{\rm{T}} = E_{\max} = 7$ TeV, $\beta_{\rm{T}} = 2.3$, $\Gamma_{\rm{T}} = 560$ \citep{2023Sci...380.1390.}, $E_{\rm \max,o} = 10$ MeV, and $L_{\rm{M}}(t_{\rm{R}})$, $E_{\rm{P}}(t_{\rm{R}})$, $\alpha_{\rm{M}}(t_{\rm{R}})$, $\beta_{\rm{M}}(t_{\rm{R}})$ obtained from \cite{2023arXiv230301203A} and \cite{2023ApJ...949L...7F}. 
The data from \cite{2023ApJ...949L...7F} cover a longer time interval, while the data from \cite{2023arXiv230301203A} have a higher time resolution. 
Hence, we use the data from \cite{2023arXiv230301203A} to describe the pulse and the data from \cite{2023ApJ...949L...7F} for other time intervals. Here, $t_{\rm{T}} = T - T^*$, where $T^* = 226$ seconds\citep{2023Sci...380.1390.}.

The parameter $t_{\rm{T}}$ influences $\bar{\tau}$ through $R_{\rm{T}}$ and the light curve of MeV photons.
The three curves (full formula, without $\delta t$ and $E_{\rm{P}}^* = E_{\rm{P}}$) exhibit obvious bends due to the light curve, while the remaining curves (constant $L_{\rm{M}}$ and simple), not considering the effect of light curve, are almost straight lines. 
We see that $R_{\rm{T}}$ has a larger impact on $ \bar{\tau}$ in the case of full formula (corresponding to black thick line in figure \ref{fig:general}) and $E_{\rm{P}}^* = E_{\rm{P}}$ (corresponding to purple dotted line in figure \ref{fig:general}), but the light curve of MeV photons has a larger impact on $ \bar{\tau}$ in the case of without $\delta t$(corresponding to green dashed dotted line in figure \ref{fig:general}). 
From equation \eqref{eq:e_min}, we find that $E_{\min}$ is much smaller than $E_{\rm{P}}$, indicating that low-energy MeV photons contribute more to the optical depth. In the simple model(corresponding to red thin line in figure \ref{fig:general}), we choose the low-energy power-law index as the spectral index. The difference between the simple models and the full formula is not significant.

The results obtained using constant luminosity (corresponding to orange dashed line in figure \ref{fig:general}) are smaller than the full formula at the beginning and larger than the full formula towards the end. This is because we used the average value of the entire pulse, and the number density of MeV photons is lower than the number density at the peak but higher than the number density after the pulse.

The results obtained using $E_{\rm{P}}^* = E_{\rm{P}}$ are much larger than the full formula because $E_{\rm{P}} > E_{\rm{P}}^*$ at every time, and the optical depth $\tau \propto {E_{\rm{P}}^*}^{-2}$ due to equation \eqref{eq:simple_tua}. The significant difference between the two curves shows that the correction for the photon number density with the power-law index of the spectrum is important.

The difference between the full formula and the method that neglect $\delta t$ correction indicates the importance of this correction. The two curves are similar in shape. To better illustrate the difference between the two, we need to make some approximations. The photon density decays with the square of the radius, so the minimum radius $R_{\rm{T}}$ has the maximum impact on the optical depth. As mentioned in the previous section, the maximum angle $\theta_{\rm{T}} = 1/\Gamma_{\rm{T}}$ has the greatest impact on the light depth. Hence, $\delta t \simeq \frac{1}{2}{\theta_{\rm{T}}}^2R_{\rm{T}}/c = t_{\rm{T}}$, in other words, the complete form is delayed by approximately $t_{\rm{T}}$ time. On the other hand, the optical depth of the full formula is influenced by the entire light variation curve, so the curve is smoother. The peak in the two curves is caused by low $E_{\rm{P}}$, leading to a large number density of MeV photons.

Next, we set $ \bar{\tau} = 1$ to obtain the minimal Lorentz factor. 
The results are shown in figure \ref{fig:general} and a selected time of $t_{\rm T}$ are shown in table \ref{tab2}.
Our minimal Lorentz factor is below 300 for $t_{\rm T} = 5 {\rm s}$, which is the earliest time for TeV photons are certainly optically thin seeing from figure 3 of \citep{2023Sci...380.1390.}.
This value is much smaller than the single zone model,
and is consistent with the Lorentz factor estimated by the afterglow model, where the Lorentz factor of the early TeV emitting region being around 560 \citep{2023Sci...380.1390.}. 
According to equations (\ref{eq:simple_tua}) and (\ref{eq:simple_gamma}), the optical depth $ \bar{\tau}$ is more sensitive to changes in luminosity compared to the minimum Lorentz factor $\Gamma_{T,min}$. Hence, the difference of optical depth between the full formula, constant $L_{\rm{M}}$, and the version without $\delta t$ correction is larger than the minimum Lorentz factor.
One can see the minimal Lorentz factor decreases with time. The reason is that with time increases, the radius of the TeV emitting region increases, and consequently, the optical depth decreases.

\begin{table}
\begin{center}
\begin{tabular}{ccccccc}
   \toprule
{$t_{\rm T}$ (s)} & {full formula} & {constant $L_M$} & {without $\delta t$} & {simple model} & {$E_P^* = E_P$} \\
 \midrule
5 & 297.92 & 266.92 & 308.42 & 192.88 & 373.35 \\
10 & 162.93 & 222.37 & 154.26 & 158.40 & 201.50 \\
15 & 139.31 & 199.67 & 138.88 & 141.17 & 174.89 \\
20 & 113.06 & 184.90 & 113.00 & 130.09 & 137.51 \\
\bottomrule
\end{tabular}
\caption{The minimum Lorentz factor as described by different methods in the two-zone scenario at different time. The code can be found at \url{https://github.com/qilunuo9/two-zone_model}.}
  \label{tab2}
\end{center}
\end{table}

\begin{figure}[ht!]
\plotone{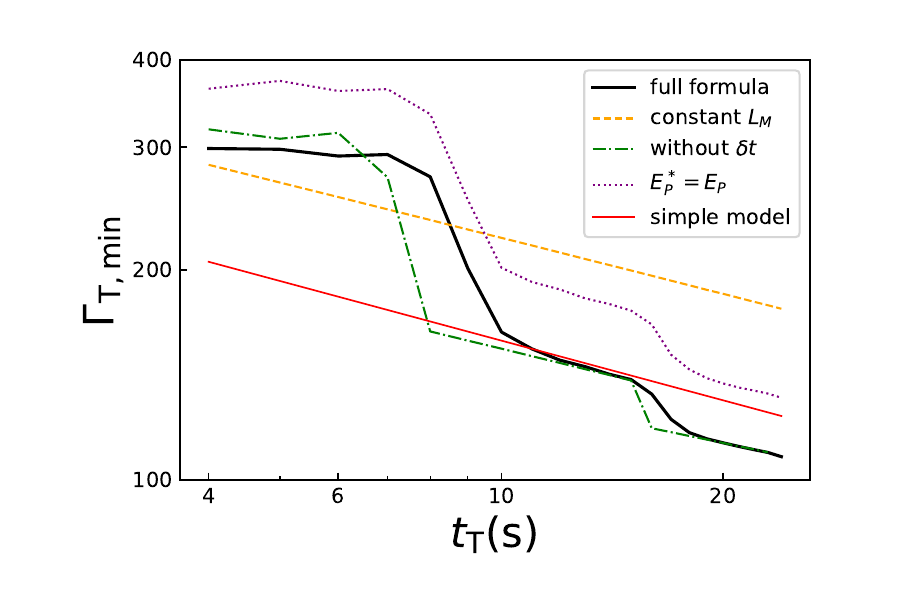}
\caption{Minimal Lorentz factor $\Gamma_{\rm T,min}$ with $E_{\rm{T}} = 7 \, {\rm TeV}$ as a function of $t_{\rm{T}}$. 
Different lines and line style correspond to different method. The black thick solid line is got by full formula. 
We neglect change of luminosity, power law indexes and peak energy when we get orange dashed one. 
We neglect $t_{\rm{R}}$, only consider change of $t_{\rm{T}}$ when we get the green dashed dotted one. 
The red thin solid one is got by the simple method.
The purple dotted one omits the correction for photon number density with power law index of spectrum, that is to say, $E_{\rm{P}}^* = E_{\rm{P}}$.}
\label{fig:general2}
\end{figure}

\section{conclusion and discussion}
\label{sec:con}
We discussed the constraints on the jet Lorentz factor for the extremely bright gamma-ray burst GRB 221009A, based on the observation of very high-energy (TeV) photons during the prompt phase.
The detection of TeV photons during the prompt phase is significant because typically, TeV photons should collide with lower energy photons and get absorbed. The fact that we observe TeV photons suggests a very high Lorentz factor that allows the TeV photons to escape without being significantly absorbed.
To estimate the optical depth for TeV photons, we used a two-zone model, where the TeV photons come from the external shock region, while lower energy photons come from the internal prompt emission region. It is crucial to correct for the MeV photon spectrum evolution and light curve to obtain an accurate estimate of the optical depth.
The most stringent constraint obtained from our calculations is that the minimum Lorentz factor for the TeV photon emitted at 5 s is about 300. This is much lower than the estimate from the single-zone model, which highlights the importance of considering the two-zone model with spectral and light curve corrections for accurate Lorentz factor estimates.
Our two-zone model with spectral and light curve corrections provides reasonable Lorentz factor estimates that are consistent with afterglow modeling. 
Notice that the constraint is just for the TeV emitting zone.


In this study, we required the gamma-ray burst to be sufficiently bright to provide an adequate number of photons for accurately fitting the power-law photon index in each time interval. This ensured that our analysis was based on robust data.
We made certain assumptions to simplify the modeling process. First, we assumed that the emission radius of MeV photons is much smaller than that of TeV photons, making them nearly indistinguishable from being emitted directly from the central engine.
Second, we assumed that the emission angle of MeV photons is larger than that of TeV photons.
Although 7 TeV photons are only observed during a certain time, we made the reasonable assumption that they remain optically thin throughout the entire period for TeV photons. This assumption is supported by the fact that $\Gamma_{\rm T, \min}$, which represents the minimal Lorentz factor, does not sensitively depend on the energy of the TeV photons, as shown in equation (\ref{eq:simple_tua}).

\cite{2023arXiv230714113D} proposed that both TeV and MeV photons originate from the same radius in the ICMART model. They also considered $\gamma-\gamma$ annihilation but reduce the optical depth by assuming that a gamma-ray burst is composed of multiple sub-pulses. Additionally, they assumed a larger radiation radius, which results in optical depths less than one and allows for reasonable Lorentz factors.

In the future, as more gamma-ray bursts with photon power-law indices that can be fitted in segmented time intervals are observed, we can efficiently estimate their minimum Lorentz factors using existing codes. Additionally, with the continuous improvement of instrument precision, the required brightness threshold for gamma-ray bursts is expected to decrease, leading to more frequent observations of such bursts. 

\begin{acknowledgments}
We thank for the helpful discussions with Aming Chen, Kai Wang, Weihua Lei, Hao Wang, Jun-Yi Shen, Yuan-Yuan Zuo, Lin Zhou, and Rui-Hang Dong.
The schematic figure was plotted with GeoGebra.
The English was polished with ChatGPT.
This work is supported by the National Natural Science Foundation of China (grant No. U1931203).
\end{acknowledgments}

\bibliography{LF-2-zone}{}
\bibliographystyle{aasjournal}


\end{document}